# Fission involves a new state of nuclear matter

C. YTHIER, S. HACHEM and G. MOUZE

*Faculté des Sciences, 06108 Nice cedex 2, France*

**Abstract**- The rearrangement step of nuclear fission occurs within 0.17 yoctosecond, in a new state of nuclear matter characterized by the formation of closed shells of nucleons. The determination of its lifetime is now based on the prompt neutron emission law. The width of isotopic distributions measures the uncertainty in the neutron number of the fragments. Magic mass numbers, 82 and 126, play a major role in the mass distributions. Arguments are presented in favour of an all-neutron state. The boson field responsible for the new collective interaction has to be searched for.

**Introduction.** - An overall picture of our model of binary nuclear fission was recently given by R.A. Ricci in Europhysics News [1]. But seeing that F. Gönnenwein does not believe [2] that fission occurs within $1.7 \cdot 10^{-25}$s, we will first try to justify that this holds for all the fissioning systems considered by J. Terrell in his work on prompt neutron emission [3], and then we will try to show that the idea of closed "nucleon shells" was already contained in another paper by J. Terrell [4] and can explain the mass distributions of binary fission.

**Fission occurs within 0.17 yoctosecond-** In 1957, J. Terrell showed that the probability $P(\nu)$ of emitting $\nu$ neutrons per fission, represented as function of the difference $(\nu - \bar{\nu})$, where $\bar{\nu}$ is the average number of neutrons emitted per fission, is a Gaussian curve having a $\sigma$ - parameter of 1.08, or a full-width–at-half-maximum of 2.538 neutrons. Indeed, the data obtained from the following spontaneously fissioning nuclei, $^{238,240,242}$Pu, $^{242,244}$Cm and $^{252}$Cf and from $^{233,235}$U and $^{239}$Pu irradiated with 80 keV neutrons, were perfectly fitted by such a curve, as demonstrated by his figure 4 [3], reproduced in many textbooks, e.g. in [5]. J. Terrell tried to explain this situation as resulting from large variances in total kinetic energy and hence in excitation energy of the fragments.

In the nineties, the study, in particular by J.H. Hamilton et al. [6,7], using coincidence methods and detectors such as Gammasphere, of the prompt gamma rays emitted by neutron-rich fission fragments, revealed that the isotopic distributions of fission fragments can be represented by Gaussian curves, and J.L. Durell [8] pointed out that all isotopic distributions encountered in fission have exactly the same width as that found by J. Terrell for the $P(\nu)$ distribution; as an example, J.L. Durell showed that all the isotopic distributions of Zr fragments associated in γ-γ coincidences with each of the Ba fragments from $^{142}$Ba to $^{148}$Ba, observed in the spontaneous fission of $^{248}$Cm, have a width of ~ 2.54 mass units (u) . J.L.Durell proposed the same explanation as Terrell.

Let us look at the particular Zr-isotopic distribution which corresponds to coincidences with $^{148}$Ba, displayed e.g. in ref.[8]. As a consequence of matter conservation, this distribution cannot extend beyond N=60, i.e. beyond $^{100}$Zr, for the $^{248}$Cm fissioning system. And, for the same reason, no prompt neutron can have been emitted at N=60, as noted by J.L. Durell himself: Isotopic distributions of fission fragments teach us how these fragments are formed at the same time that prompt neutron emission occurs.

Another observation can be made in this distribution. The abscissa $\bar{N}$ of the maximum of the Gaussian distribution is the *most probable value of the "neutron number N"* of the various Zr isotopes which can be formed in this particular coincidence. And, as expected for a random phenomenon, $\bar{N}$ is non-integer. Indeed, a value $\bar{N}$ ~ 58.4 has been found by J.L. Durell in the coincidences with $^{148}$Ba, whereas he found $\bar{N}$ ~ 61.8 in the $^{142}$Ba case.

Moreover it may be asked why the width of the Gaussian distributions has a constant value ~ 2.54 u , in so many different experiments, and for so many different fissioning systems: This situation clearly indicates that *prompt neutron emission and fragment formation* both occur within an extremely short time interval Δt.

In our opinion, due to the brevity of the fission reaction, the *energy-time uncertainty relation* ΔE. Δt = ℏ has to play a major role in its description. To the uncertainty ΔE corresponds an uncertainty ΔA = ΔE/$c^2$, and to ΔA correspond uncertainties ΔN in N and ΔZ in Z. And the finite and constant value of the width ΔN can be interpreted as the "uncertainty in the neutron number N" of the fragment. With ΔN = (N/A) ΔA, and with a value 1.6449 for the mean value of A/N in the fissioning systems taken into consideration by J. Terrell, we get the following value for $\overline{\Delta t}$, if $\overline{\Delta N}$ =2.538 u:

$$\overline{\Delta t} = \hbar/c^2 \; \overline{(A/N)} \; \overline{\Delta N} = 1.696 \; 10^{-25} \text{ s} \qquad (1)$$

This mean value of the reaction time of the fission reaction for the light actinide nuclei is now based on the considerable work made in the fifties for determining $\bar{v}$ and P(v), i.e. on "data which are of the utmost practical importance in the application of the chain reaction in reactors and explosives" [9].

A value Δt = 1.77 10$^{-25}$s , quoted in [1], had been announced in 1993 by G. Mouze and C. Ythier [10].This value was based only on cold fission data concerning the reaction $^{235}$U + n$_{th}$ at the highest total kinetic energy [11]. A similar value was recently deduced from other old cold fission experiments [12].

To each primary fission fragment are attached, as a kind of trademark, uncertainties $\Delta A$, $\Delta N$ and $\Delta Z$. $\overline{\Delta Z}$ is given by the difference $\overline{\Delta A} - \overline{\Delta N} = 1.637$ proton; it is the modern expression of the distribution law of the charge in fission.

It must be pointed out that a reaction time of 0.17 yoctosecond is not surprising for a reaction occurring within an atomic nucleus. If the reaction time of a chemical reaction can be defined as the ratio of range $\ell$ of interaction to velocity of propagation of the interaction, its smallest value is $\ell/c$, i.e., for a reaction occurring between H and Cl in a HCl molecule, ~0.43 $10^{-18}$ s, with $\ell \sim 1.27455\ 10^{-10}$ m [13]. And for a reaction occurring within the most external valence shells of a nucleus, $\ell$ can be about $10^7$ times smaller, and $\Delta t$ can be of the order of 1 $10^{-25}$ s. Interestingly, the value $\Delta t = 0.17$ ys corresponds to a range of ~5.09 $10^{-17}$ m.

**The closure of nucleon shells at A = 82 and A =126.-** During the time interval $\Delta t$, an energy $\Delta E \cong 3.8$ GeV is at disposal of the fissioning system. This corresponds to a temperature $T = \Delta E/k = 4.5\ 10^{13}$ K. One may speak of extreme conditions.

In 2008 Mouze and Ythier [14] suggested that in so extreme conditions a new state of nuclear matter could be created, in which any distinction between proton and neutron seems to have disappeared. And they showed that the mass distributions of asymmetric fission, and even those of symmetric fission, could be explained if one assumes that "nucleons", instead of differentiated protons and neutrons, form, as a consequence of the spin-orbit coupling, closed shells at magic "mass numbers" 82 and 126.

Indeed, if one assumes, as proposed by Ythier and Mouze at the Karlsruhe Symposium [15], that a fissioning system, such as $^{235}$U + $n_{th}$, can "clusterize" into a dinuclear system $^{208}$Pb + $^{28}$Ne, with a great energy- release, the possibility of a collision of core and cluster cannot be excluded. But what can happen in such an internal collision? It cannot simply be the capture, by the cluster, of 76 valence nucleons, or less, of the deep-lying doubly magic $^{132}$Sn core, as initially proposed [16]: the hypothesis of the creation of an A = 126 "nucleon core" in $^{208}$Pb leads to a better description of the mass distributions occurring in fission.

In the asymmetric fission of $^{235}$U + $n_{th}$, for example, the transfer of the 82 valence nucleons released by $^{208}$Pb allows the formation of an A = 82 "nucleon core" around the cluster, of mass number $A_{cl}$ (equal to 28 in this case). The mass number of the light fragment, $A_L$, is also comprised between $A_L^{MIN} = 82$ and $A_L^{MAX} = A_{cl} + 82$, i.e. 110, whereas $A_H$ is comprised between $A_H^{MIN} = 126$ and $A_H^{MAX} = 126 - [82 - (82-A_{cl})]$ = 126 + $A_{cl}$, i.e. 154. The width of the region of appreciable yield, $\Delta A = A_{MAX} - A_{MIN}$, is the same for light and heavy fragment, and equal to $A_{cl}$ (first Hachem rule [17]), i.e. $\Delta A = 28$ u.

In the symmetric fission of nuclei heavier than $^{252}$Cf, e.g. in that of $^{258}$Fm (or $^{208}$Pb-$^{50}$Ar), the formation of an A = 126 "nucleon core" becomes possible even in the light fragment. This formation explains the narrow width of the distributions. Indeed, $A_L^{MIN}$ = 126, $A_L^{MAX}$ = $A_{cl}$ + 82 = 132, $A_H^{MIN}$ = 126, $A_H^{MAX}$ = 126 − [82 − (126−$A_{cl}$)] = $A_{cl}$ +82 =132. Thus  ΔA = ($A_{cl}$ − 44) u = 6 u (second Hachem rule)[18] for the now single peak of the distribution.

The width of 8 u found experimentally [19] can be justified by the uncertainty in A [20]; in fact $\Delta A_{exp}$ ~ ($A_{cl}$ -44) + 2 u.

The idea of closed nucleon shells can find its justification in observations made by J. Terrell in 1962 [4]. First, he reported that fission fragments with A = 82 and A = 126 do not emit prompt neutrons, and that the prompt neutron yield increases linearly above these A- values; more precisely, the number of prompt neutrons emitted by the light fragment can be represented by ν(L) = 0.08 ($A_L$ -82) and that emitted by the heavy fragment by ν(H) = 0.10 ($A_H$ − 126). Moreover, concerning the lower limits of the mass distributions, he made the important observation that *"asymmetric fission seems to be characterized by the relation $A_L$ > 82, $A_H$ > 128: these limits seem to define quite accurately the region of appreciable yield. They also seem to be the point at which neutron yield nearly vanishes"*. He noted that $A_H$ = 126 has to be preferred to A = 128, because 126 *"gives a better linear representation of ν(H)"*.

**Could the nucleon phase be an all-neutron state ?.-** The "nucleons" of this phase deserve to be called nucleons since mass-numbers, rather than proton- and neutron-numbers, without doubt play a major role. But what are these nucleons? The survival, in this phase, of the magic numbers 82 and 126 clearly suggests that *the "spin- orbit coupling* still plays a role, and consequently that these particles still have a *spin*. But it may be asked: Are they all protons, or all neutrons? We present at least two arguments in favour of a neutron state, i.e. in favour of a disappearance of the charge. First we observed that in ordinary nuclear matter, it is *the Coulomb potential of the protons* that is responsible for the particular series of energy states of the proton phase. Indeed, the series of states of the neutron phase result from the sole spin- orbit coupling. Thus, the disappearance of the protons, because they are changed into neutrons, would lead to a pure neutron phase, and this justifies that magic "mass numbers" such as 82 and 126 intervene. Secondly, an all-neutron state during the rearrangement step seems necessary for the formation of fragments in conditions in which their neutron-number is affected by a considerable uncertainty. Moreover at the end of this step, as protons are present again, the ratio N/Z is still so great that a series of beta decays is necessary in order to get the final much smaller value.

**Could the creation of the new phase be related to an interaction with W$^+$ and W$^-$-boson fields?** – It may be asked whether the all-neutron state can be explained by the known interactions encountered in high-energy physics, e.g. by the interaction with the $W^\pm$-boson field. It is well known that the $W^-$ boson field is involved in the $\beta^-$ decay of the fission products. There, a d-quark becomes changed into a u- quark at each step of the decay chain. If the charge of Z protons of the fissioning system disappears, and then reappears, it means that now a number Z of u-quarks are changed into a number Z of d- quarks, which then reappear, changed into a number Z of u – quarks.

But can the interaction with the $W^-$ fields have such a collective character that, instead of only one, in fact Z quarks can be changed? Moreover, what happens to the leptons involved in these quark-changes? Is it necessary to assume that these changes occur within a so short time interval that no lepton is emitted? In fact, the lifetime of the new phase is equal to $1.697 \times 10^{-25}$s: it is remarkably shorter than that of the $W^\pm$-boson, i.e. $3.074 \times 10^{-25}$ s , and even shorter than that of the $Z^0$- boson, i.e. $2.6378 \times 10^{-25}$ s [21]; maybe, such a lifetime is short enough for explaining that no one lepton is observed. Moreover, this extremely short lifetime suggests that it could correspond to *a very heavy* intermediary new particle. Indeed, one observes that the ratio $R = M/\Gamma$ of mass and energy-width is almost constant for the $W^\pm$ and $Z^0$ – bosons, and equal to 35.74 and 36.548 respectively. With $\Gamma$ equal to $\hbar/\Delta t = 6.582 \times 10^{-25}$ GeVs/ $1.697 \times 10^{-25}$ s, i.e. to 3.878 GeV, for the nucleon phase, a mass equal to $3.878 \times 36.146 = 140.174$ GeV/$c^2$, or even greater, could be predicted for the intermediary particle; but, why not a mass greater than $M(2W) = 160.8$ GeV/$c^2$, the mass of the $W^+$ -$W^-$ boson *pair* ? An interaction of the fissioning system with $W^+$ *and* $W^-$ fields could so be justified.

**Conclusion.-** One sees that the extreme brevity of the fission reaction can no more be contested. *Fission occurs more than thousand times faster than believed since seventy years.*

Symmetric as well as asymmetric mass distributions can now be explained. This explanation reveals that *spin-orbit coupling is a fundamental property of nuclear matter*, since it survives in extreme conditions.

The fact that the charge of all protons disappears and reappears in the ephemeral new phase strongly suggests that *they interact with the $W^+$ and $W^-$ boson fields*. But what do we really know about the true nature of the charge [22] ?